\documentclass[aps,prl,twocolumn,showpacs,amsmath,amssymb]{revtex4-1}

\usepackage{graphicx}
\usepackage{color}

\definecolor{Red}{rgb}{1.0,0.0,0.0}

\def\be#1{\begin{equation}\label{#1}}
\def\ee{\end{equation}}

\begin{document}

\title{Gauge Field Emergence from Kalb-Ramond Localization}

\author{G. Alencar${}^a$}
\email{geova@fisica.ufc.br}
\author{R. R. Landim${}^a$}
\email{rrlandim@gmail.com}
\author{M. O. Tahim${}^{b,}$}
\email{makarius.tahim@uece.br}
\author{R.N. Costa Filho${}^a$}
\email{rai@fisica.ufc.br}
\affiliation{${}^a$Departamento de F\'{\i}sica, Universidade Federal
do Cear\'a, 60451-970 Fortaleza, Cear\'a, Brazil}
\affiliation{${}^b$Universidade Estadual do Cear\'a, Faculdade de Educa\c c\~ao, Ci\^encias e Letras do Sert\~ao Central - 
R. Epit\'acio Pessoa, 2554, 63.900-000  Quixad\'{a}, Cear\'{a},  Brazil.}

\date{\today}

\begin{abstract}

A new mechanism, valid for any smooth version of the Randall-Sundrum model, of getting localized massless vector field on the brane is described here. This is obtained by dimensional reduction of a five dimension massive two form, or Kalb-Ramond field, giving a Kalb-Ramond and an emergent vector field in four dimensions. A geometrical coupling with the Ricci scalar is proposed and the coupling constant is fixed such that the components of the fields are localized. The solution is obtained by decomposing the fields in  transversal and longitudinal parts and showing that this give decoupled equations of motion for the transverse vector and KR fields in four dimensions. We also prove some identities satisfied by the transverse components of the fields. With this is possible to fix the coupling constant in a way that a localized zero mode for both components on the brane is obtained. Then, all the above results are generalized to the massive $p-$form field. It is also shown that in general an effective $p$ and $(p-1)-$forms can not be localized on the brane and we have to sort one of them to localize. Therefore, we can not have a vector and a scalar field localized by dimensional reduction of the five dimensional vector field. In fact we find the expression $p=(d-1)/2$ which determines what forms will give rise to both fields localized. For $D=5$, as expected, this is valid only for the KR field.
  
\end{abstract}

\maketitle

\section{Introduction}

In Kaluza-Klein models with extra dimensions (string theory and others) the most basic tool is the decomposition of fields depending on the dimensions they are embedded and its tensorial characteristics. For example, working in $D=5$ and taking the important field as $g_{\mu\nu}$, the dimensional reduction to $D=4$ will give us again a four dimensional gravitational field, a vector field, and a scalar field (the dilaton) as dynamical actors. Enlarging the number of extra dimensions we can add Yang-Mills fields in the procedure of dimensional reduction to $D=4$ \cite{Bailin:1987jd}. The same can be made to $p-$form fields. For fermion fields there is the specific procedure to obtain in lower dimensions several kinds of fermionic fields (chiral or not, real or not). We present in this work a similar procedure that can be applied to localize $p-$form fields in the  Randall-Sundrum scenario of extra dimensions \cite{Randall:1999vf,Randall:1999ee}. Interestingly, the results are similar to the fermion case and by dimensional reduction we generally have that some components of the lower dimensional fields are not localized.  It is important to mention that this procedure actually provides a new mechanism to localize gauge vector fields: from a Kalb-Ramond field in $D=5$ we can obtain the $4D$ Kalb-Ramond and an additional localized vector field. We can think the gauge field emerges in this mechanism.
 
The problem of gauge form field localization in several brane world scenarios has been studied along the last years. This is a necessary step to walk along since our four dimensional space-time presents us a propagating vector field, despite more possible signals which can be interpreted as coming from other tensor gauge fields. In this sense, it is already understood how to localize the zero modes of gravity and scalar fields \cite{Randall:1999vf,Bajc:1999mh} in a positive tension brane. However, the conformal invariance of the basic vector model fall into serious problems for building a realistic model because the localization method gives no result. This problem has been approached in many ways. Some authors have introduced a dilaton coupling in order to solve it \cite{Kehagias:2000au} and other propose that a strongly coupled gauge theory in five dimensions can generate a massless photon in the brane \cite{Dvali:1996xe}. Modifications of the model considering spherical branes, multiple branes or induced branes can be found in \cite{Gogberashvili:1998iu,Jardim:2011gg,Jardim:2013jy,Akama:2011wi,Kaloper:1999et,Kaloper:2004cy,Alencar:2013jm,Akama:1978pg,Akama:2014fra,Akama:2013tua}.

Beyond the gauge field (one form) other forms can be considered. In five dimensions we can have yet the two, three, four and five forms. In $D-$dimensions we can in fact think about the existence of any $p\leq D$. However, as we will see, they can be considered in a unified way. The analysis of localizability of the form fields has been considered in \cite{Kaloper:2000xa} where it has been shown that in $D-$dimensions only the forms with $p<(D-3)/2$ can be localized. However, it is well known that in the absence of a topological obstruction, the field strength of a $p-$form is dual to the the $(D-p-2)-$form \cite{Duff:1980qv}. Using this the authors in \cite{Duff:2000se} found that also for $p>(D-1)/2$, the fields are localized. It is important to point that in the model proposed here the Hodge Duality is not valid since we consider mass terms in the action that break the duality. Beyond the zero mode localization the resonances of $p-$forms has also been studied \cite{Alencar:2010vk,Alencar:2010hs,Landim:2010pq,Alencar:2010mi,Fu:2012sa}.

Another interesting point of view is related to models where membranes are smoothed out by topological defects
 \cite{Bazeia:2005hu,Liu:2009ve,Zhao:2009ja,Liang:2009zzf,Zhao:2010mk,Zhao:2011hg,Landim:2011ki,Landim:2011ts,Alencar:2012en}. The advantage of these models is that the $\delta$-function singularities generated by the brane in the RS scenario are eliminated. This kind of generalization also provides methods for finding analytical solutions \cite{Cvetic:2008gu,Landim:2013dja}. This is a nice characteristic if we want to put forward the idea of considering a geometrical coupling with the Ricci scalar. The Ricci scalar can inform about possible space-time singularities and, as we want avoid them, such a coupling is natural in this sense. We therefore consider this kind of coupling with the gauge field, the Kalb-Ramond field and $p$-form fields in models with smooth membranes. This kind of coupling has its origins in the DGP model and its consequences \cite{Dvali:2000hr}. One of its consequences is a model of (quasi) localization of gauge fields \cite{Dvali:2000rx} where the membrane is described by a delta function, i.e., a singular place that can be understood using the Ricci scalar: in fact we  can get that function as coming from a smooth model. The Ricci scalar, when we make the limit to the RS model, give rise to a delta function and explain the geometrical coupling with the membrane.

Other studies using a topological mass term in the bulk were introduced, but without giving a massless photon in the brane\cite{Oda:2001ux}.  Most of these models introduce other fields or nonlinearities to the gauge field \cite{Chumbes:2011zt}. As a way to circumvent this, the authors in \cite{Ghoroku:2001zu} introduced in the action, beyond the usual field strength ($Y_{MN}=\partial_{[M} X_{N]}$), a mass term in five dimensions and a coupling with the brane  given by $(M^2 +c\delta(z))G^{MN}X_M X_N$, where $X_M$ is the vector gauge field. This gives a localized massless photon. In this model the localization is obtained only for some values of the parameter $c$ and for a range in $M$. It is important to note that in this case the gauge symmetry is lost due the existence of a mass term but is recovered in the effective action of the zero mode. In this context, a model has been proposed in which the two couplings are replaced by a coupling with the Ricci scalar\cite{Zhao:2014iqa}. This is a very a natural way if we want to consider smooth version of RS model. For obtaining their results the authors of \cite{Zhao:2014iqa} used the particular configuration of fields $\partial_{\mu}A^{\mu}=A_5=0$. This is the same gauge used in the massless case. However, here we have a mass term and the gauge symmetry is lost. Therefore, the result obtained by them is not generally valid. A solution to this problem was found by the present authors in \cite{Alencar:2014moa}. We show there that the choice $\partial_{\mu}A^{\mu}=A_5=0$, yet being valid as a particular solution, is unnecessary. We show that upon dimensional reduction of the five dimensional vector field ($A_{M}$) we get decoupled equations for the scalar ($A_5$) and the transverse vector ($A_{\mu}$) fields  in four dimensions.  For this we prove some identities satisfied by the transverse component of the field $A_{\mu}$.  Then we obtain that we just can localize the zero mode of the $A_{\mu}$ or of the scalar field. 

In the present manuscript we consider the same procedure to the two form field, which by dimensional reduction gives us a two and an one form fields in four dimensions. In this case we obtain that both fields are simultaneously localized on the four brane.   Therefore, as commented before, we find that we can have to different situations: upon dimensional reduction some components of the lower dimensional fields are not localized. A special case happens for the KR field in $D=5$. To have a better understanding of this we  generalize our results to higher dimensions and consider $p-$forms fields on it.  We find that  for each space-time dimension $D$ we can have just one higher dimensional $p-$form which provides both components of lower dimensional form fields localized. In fact we find a relation, given by $p=(D-1)/2$, where this is valid.

The paper is organized as follows. In section two we review the results for the one form gauge field. In section three we study the generalization for the Kalb-Ramond, or two form field. After considering similar decomposition of the field we show that they are decoupled. By dimensional reduction it is also shown that we can localize both, the gauge and the Kalb-Ramond fields in four dimensions. In section four we generalize all the results to the $p$-form case.

\section{The One Form Case}
Here we must review the results found by the authors in a previous work \cite{Alencar:2014moa}. The geometrical coupling is proposed with action
\begin{eqnarray}\label{geometrical}
&S_{1}=-\int d^{5}X\sqrt{-g}( \frac{1}{4}g^{MN}g^{PQ}Y_{MP}Y_{NQ} \nonumber \\
&-\frac{\gamma_{1} }{2}\int d^5x\sqrt{-g} Rg^{MN}X_{M}X_{N},
\end{eqnarray}
where $ds^2=e^{2A(z)}(dx_\mu dx^\mu+dz^2)$. The equations of motion are
\begin{equation}\label{eqmotion}
\partial_{M}(\sqrt{-g}g^{MO}g^{NP}Y_{OP})=-\gamma_{1}\sqrt{-g}Rg^{NP}X_{P},
\end{equation}
and from the antisymmetry of Eq.  (\ref{eqmotion}) obtain the transverse condition $\partial_{N}(\sqrt{-g}RX^{N})=0$. Then split the field in two parts $X^{\mu}=X_{L}^{\mu}+X_{T}^{\mu}$,
where $L$ stands for longitudinal and $T$ stands for transversal with $ X_{T}^{\mu}=(\delta_{\nu}^{\mu}-\frac{\partial^{\mu}\partial_{\nu}}{\Box})X^{\nu}$ and $X_{L}^{\mu}=\frac{\partial^{\mu}\partial_{\nu}}{\Box}X^{\nu}$.
With this, Eq.  (\ref{eqmotion}) can be divided in two. For $N=5$ 
\begin{equation}
\label{M=5}
\partial_{\mu}Y^{\mu5}+\gamma_{1} e^{2A}R\Phi=0
\end{equation}
where $\Phi\equiv X_5$ and for $N=\nu$ we get
\begin{equation}
\label{M=mu}
e^{A}\Box X_{T}^{\nu}+(e^{A}\partial X_{T}^{\nu})'+\gamma_{1} e^{3A}RX_{T}^{\nu}+(e^{A}Y_{L}^{5\mu})'+\gamma_{1} e^{3A}RX_{L}^{\nu}=0,
\end{equation}
where the prime means a $z$ derivative, and all lower dimensional index will be contracted with $\eta^{\mu\nu}$. Yet form our transversality condition we get
 \begin{equation}
\label{transverserelation}
e^{3A}R\partial_{\mu}X^{\mu}=-(e^{3A}R\Phi)'
\end{equation}
and using the previous definition and $Y_{L}^{5\mu}\equiv  X_{L}^{'\mu} -\partial^{\mu}\Phi$ we can show the following identities
\begin{equation}
\label{identities}
\partial_{\mu}Y^{\mu\nu}=\Box X_{T}^{\nu}; Y^{5\mu}=X_{T}^{'\mu}+Y_{L}^{5\mu}; Y_{L}^{\mu5}=\frac{\partial^{\mu}}{\Box}\partial_{\nu}Y^{\nu5}.
\end{equation}
Using now (\ref{M=5}), (\ref{transverserelation}) and (\ref{identities})  we get
\[
(e^{A}Y_{L}^{\mu5})'=-\gamma_{1}\frac{\partial^{\mu}}{\Box}(e^{3A}R\Phi)'=-\gamma_{1} e^{3A}RX_{L}^{\nu},
\]
and finally obtain from Eq.  (\ref{M=mu}) the equation for the transverse part of the gauge field
\[
e^{A}\Box X_{T}^{\nu}+(e^{A}\partial X_{T}^{\nu})'+\gamma_{1} e^{3A}RX_{T}^{\nu}=0.
\]

Finally separating the $z$ dependence like $X_{T}^\mu=\tilde{X}_{T}^\mu\tilde{\psi}(z)$, using $R=-4(2A''+3A'^{2})e^{-2A}$ and performing the transformation $\tilde{\psi}=e^{-\frac{A}{2}}\psi$ we get the desired Schr\"odinger equation with potential
 \begin{equation}
\label{effectivepotential}
U=(\frac{1}{4}+12\gamma_{1})A'^{2}+(\frac{1}{2}+8\gamma_{1})A''
\end{equation}
which is localized for $\gamma_{1}=1/16$ with solution $e^{A}$. Here we correct a misprint of Ref. \cite{Alencar:2014moa} where we gave the solution $e^{A/2}$. For the scalar field we must be careful since we have
\[
\Box\Phi-(\partial_{\mu}A^{\mu})'-\gamma_{1}Re^{2A}\Phi=0.
\]

Performing the separation of variables $\Phi=\Psi(z)\phi(x)$, defining $\Psi=(e^{3A}R)^{-1/2}\psi$, using Eq.  (\ref{transverserelation}) and after some manipulations we get a Schr\"odinger equation for the massive mode of the scalar field with potential given by \cite{Alencar:2014moa} 
\[
U=\frac{1}{4}(3A'+(\ln R)')^{2}-\frac{1}{2}(3A''+(\ln R)'')+\gamma_{1}Re^{2A}.
\]
With this potential we see that the zero mode of the scalar field solution is localized for $\gamma_{1}=9/16$. This shows us that we cannot have both fields localized.

 \section{The Kalb-Ramond Field Case}
In this section we use the same approach as before in order to try to
localize the zero mode of the Kalb-Ramond field. Upon dimensional
reduction of the KR field we are left with to kinds of terms, namely
a Kalb-Ramond in four dimensions $B_{\mu\nu}$ and a vector field
$B_{\mu5}$. We must remember that here we also do not have gauge symmetry
and we can not  choose $B_{5\mu}=0$. However, we can again show
that the longitudinal and transversal parts of the field decouples
and we get the desired results. The action in this case is given by 
\[
S_{2}=\int d^{5}x\sqrt{-g}\left[-\frac{1}{24}(Y_{M_{1}M_{2}M_{3}})^2-\frac{1}{4}\gamma_{2}R(X_{M_{1}M_{2}})^2\right],
\]
and the equations of motion are given by 
\begin{equation}\label{motion2form}
\frac{1}{2}\partial_{M_{1}}\left[\sqrt{-g}Y^{M_{1}M_{2}M_{3}}\right]-\gamma_{2} R\sqrt{-g}X^{M_{2}M_{3}}=0.
\end{equation}
In the above equation all the indexes are raised with $g^{MN}$.
Just like in the case of the one form field, the antisymmetry of the equation gives us the transverse condition  $\partial_{M_{1}}(R\sqrt{g}X^{M_{1}M_{2}})=0.$ 
Now we proceed to find the decoupled equations of motion. First of
all the above equation must be expanded. For $M_{2}=\mu_{2}$ and
$M_{3}=\mu_{3}$ we obtain 
\begin{equation}\label{2formnu}
 \frac{1}{2}e^{-A}\partial_{\mu_{1}}Y^{\mu_{1}\mu_{2}\mu_{3}}+(e^{-A}Y^{5\mu_{2}\mu_{3}})' -\gamma_{2} Re^{A}X^{\mu_{2}\mu_{3}}=0;
\end{equation}
and for $M_{3}=5$ we get
\begin{equation}\label{2form5}
 \frac{1}{2}\partial_{\mu_{1}}Y^{\mu_{1}\mu_{2}5}-\gamma_{2} Re^{2A}X^{\mu_{2}}=0.
\end{equation}

The transverse equation, differently from the vector case, will give rise to two equations. For $M_{4}=5$ we get $\partial_{\mu}X^{\mu5}\equiv\partial_{\mu}X^{\mu}=0$, where
we have used the previous definitions. Therefore, we see that the transverse
condition for our vector field is naturally obtained upon dimensional
reduction. For $M_{4}=\mu_{4}$ we get 
\begin{equation}\label{transverse2formmu}
(Re^{A}X^{\mu_{4}})'+e^{A}R\partial_{\mu_{1}}X^{\mu_{1}\mu_{4}}=0.
\end{equation}
Just as in the case of the one form, here we have effective equations
that couple the Kalb-Ramond and the Vector field. Before proceeding 
to solve the equations we can further simplify them if we take the
longitudinal and transversal part of each field. As the vector field
already satisfy the transverse condition we just need to perform this
for the KR field by $X^{\mu_{1}\mu_{2}}=X_{L}^{\mu_{1}\mu_{2}}+X_{T}^{\mu_{1}\mu_{2}}$,
defined as $X_{T}^{\mu_{1}\mu_{2}}\equiv X^{\mu_{1}\mu_{2}}+\frac{1}{\Box}\partial^{[\mu_{1}}\partial_{\nu_{1}}X^{\mu_{2}]\nu_{1}}$ and  $X_{L}^{\mu_{1}\mu_{2}}\equiv -\frac{1}{\Box}\partial^{[\mu_{1}}\partial_{\nu_{1}}X^{\mu_{2}]\nu_{1}}$. Observing that 
\[
\partial_{\mu_{1}}Y^{\mu_{1}\mu_{2}\mu_{3}}=2\square X_{T}^{\mu_{2}\mu_{3}};\;\;\;\;\partial_{\mu_{1}}Y^{\mu_{1}\mu_{2}}=2\square X_{T}^{\mu_{2}},
\]
where $Y_{\mu\nu}=\partial_{[\mu}X_{\nu]}$, we see that the first
term of Eq.  (\ref{2formnu}), is already decoupled from the longitudinal part.
However, the second term is not decoupled because $Y^{5\mu\nu}=Y_{L}^{5\mu\nu}+2\partial X_{T}^{\mu\nu}$, then our equations become 
\begin{eqnarray} \nonumber
 &&e^{-A}\square X_{T}^{\mu_{2}\mu_{3}}+\partial(e^{-A}\partial X_{T}^{\mu_{2}\mu_{3}})-\gamma_{2} Re^{A}X_{T}^{\mu_{2}\mu_{3}}\\
 &&+\frac{1}{2}\partial(e^{-A}Y_{L}^{5\mu_{2}\mu_{3}})-\gamma_{2} Re^{A}X_{L}^{\mu_{2}\mu_{3}}=0 \label{eqXTXL2form}
\end{eqnarray}
and 
\begin{equation}\label{eqYLphi2form}
\frac{1}{2}\partial_{\mu_{1}}Y_{L}^{\mu_{1}\mu_{2}}-\gamma_{2} Re^{2A}X^{\mu_{2}}=0.
\end{equation}

It is clearly from Eq.  (\ref{eqXTXL2form})
that we have a coupling between the transversal part of the field,
the longitudinal part, and the gauge field. From Eq.  (\ref{eqYLphi2form})
we see that the gauge field is coupled to the longitudinal part of
the KR field. As in the case of the one form field we should expect
that we have two uncoupled effective massive equations for the gauge
fields $X_{T}^{\mu_{1}\mu_{2}}$ and $X^{\mu}$ since both satisfy
the transverse condition in four dimensions. To prove this  we use $\partial_{\mu}X^{\mu}=0$ to show that 
\[
Y_{L}^{\mu_{1}\mu_{2}5}=-\frac{1}{\Box}\partial^{[\mu_{1}}\partial_{\nu}Y^{\mu_{2}]\nu}
=2\gamma_{2}Re^{2A}\frac{\partial^{[\mu_{1}}X^{\mu_{2}]}}{\Box},
\]
where in last equality we have used Eq.  (\ref{2form5}).
Now we can use this and Eq.  (\ref{transverse2formmu}) to show
that 
\[
(e^{A}Y_{L}^{\mu_{1}\mu_{2}5})'=2\gamma_{2}Re^{A}\frac{\partial^{[\mu_{1}}\partial_{\nu_{1}}X^{\mu_{2}]\nu_{1}}}{\Box}=-2\gamma_{2}Re^{A}X_{L}^{\mu_{1}\mu_{2}}
\]
and this term cancels the longitudinal part of the mass term.
Then we get the final form of the equation of motion 
\[
e^{-A}\square X_{T}^{\mu_{1}\mu_{2}}+
(e^{-A}\partial X_{T}^{\mu_{1}\mu_{2}})'-\gamma_{2}Re^{A}X_{T}^{\mu_{1}\mu_{2}}=0.
\]
Imposing the separation of variables in the form $X_{T}^{\mu_{1}\mu_{2}}(z,x)=f(z)\tilde{X}_{T}^{\mu_{1}\mu_{2}}(x)$
we obtain the following mass equation
\[
(e^{-A}f'(z))'-\gamma_{2}Re^{A}f(z)=2m_{X}^{2}e^{-A}f(z),
\]
using the transformation $f(z)=e^{A/2}\psi(z)$ we get the standard potential, plus the correction 
\begin{eqnarray}\nonumber
U(z)&=&\left[\frac{A'^{2}}{4}-\frac{A''}{2}+\gamma_{2}Re^{2A}\right]\\ \nonumber 
&=&(\frac{1}{4}+12\gamma_{2})A'^{2}+(-\frac{1}{2}+8\gamma_{2})A''.
\end{eqnarray}
The zero mode solution is of the form $e^{bA}$ which  if plugged in the
above equation gives us $\gamma_{2}=5/16$ and we get the integrand $e^{4A}$
rendering a localized zero mode. Now we must analyze the localizability of the vector field. In order to decouple the vector field and the longitudinal part of KR field we can use Eq. (\ref{transverse2formmu}) in (\ref{eqYLphi2form}) we get
\begin{equation}
 \square X_{T}^{\mu_{2}}+[R^{-1}e^{-A}(Re^{A}X^{\mu_{2}})']' -\gamma_{2}Re^{2A}X^{\mu_{2}} = 0.
\end{equation}

Now separating the variables $X^{\mu_{1}}=u(z)\tilde{X}^{\mu_{1}}(x)$
we get the mass equation for the vector field
\begin{equation}
\left(R^{-1}e^{-A}(Re^{A}u(z))'\right)'-\gamma_{2}Re^{2A}u(z)= 2m_{1}^{2}u(z).\label{equ2}
\end{equation}
The above equation can be cast in a Schr\"odinger form by using the general transformation found in \cite{Alencar:2014moa}, or $u(z)=(Re^{A})^{1/2}\psi$. The final potential is given by 
\[
U=\frac{1}{4}(A'+(\ln R)')^{2}-\frac{1}{2}(A''+(\ln R)'')+\gamma_{2}Re^{2A}.
\]
In this way we see that for any smooth version of RS model the above potential is identical to that of the Kalb-Ramond case and we have a localized solution. In this sense, we can say that the vector field emerges in $D=4$ from the localization of the Kalb-Ramond field. In the next section it will be clear why just for the KR field in five dimensions we can have both fields localized.

\section{The $p-$form Field Case}

In this section we further develop the previous methods in order to generalize our results to the $p-$form field case in a $(D-1)$-brane. The action is given by
\begin{equation}
S_{p}=-\frac{1}{2p!}\int d^{D}x\sqrt{-g}\left[\frac{(Y_{M_{1}...M_{p+1}})^2}{(p+1)!}+\gamma_{p}R(X_{M_{2}...M_{p+1}})^2\right],
\end{equation}
where $Y_{M_{1}...M_{p+1}}=\partial_{[M_{1}}X_{M_{2}...M_{p+1}]}$. The equations of motion are given by
\begin{equation}\label{motionpform}
\frac{1}{p!} \partial_{M_{1}}[\sqrt{-g}Y^{M_{1}...M_{p+1}}]-\gamma_{p}R\sqrt{-g}X^{M_{2}...M_{p+1}}=0. 
 \end{equation}
Similarly to the one and two form case, from the above equation we get the identity
\begin{equation}\label{divpform}
 Re^{(D-p)A}\partial ^{\nu_{2}}X_{\nu_{2} N_{3}...N_{p+1}} + \left[ Re ^{(D-p)A}X_{5N_{3}...N_{p+1}}\right]' =0.
\end{equation}

Now we can obtain the equations of motion by expanding Eq.  (\ref{motionpform}).
We arrive at just two kinds of terms, where none of the
indices is $5$, giving
\begin{eqnarray}\nonumber
&&\frac{1}{p!}e^{\alpha_{p}A}\partial_{\mu_{1}}[Y^{\mu_{1}\mu_{2}...\mu_{p+1}}]+\frac{1}{p!}(e^{\alpha_{p}A}Y^{5\mu_{2}...\mu_{p+1}})'\\
&&-\gamma_{p} Re^{(\alpha_{p}+2)A}X^{\mu_{2}...\mu_{p+1}}=0,\label{pformnu}
\end{eqnarray}
with $\alpha_{p}=D-2(p+1)$.  When one of the indices is $5$ we get
\begin{equation}\label{pform5}
\frac{1}{p!}\partial_{\mu_{1}}Y^{\mu_{1}\mu_{2}...\mu_{p}5} -\gamma_{p}Re^{2A}X^{\mu_{2}...\mu_{p}5}=0. 
\end{equation}

Just like in the Kalb-Ramond case, the transverse equation (\ref{divpform}) give rise to two equations. For the index with direction $5$ we get $\partial_{\mu_{1}}X^{\mu_{1}...\mu_{p-1}5} \equiv \partial_{\mu_{1}}X^{\mu_{1}...\mu_{p-1}}=0$,
where we have used our previous definitions. Therefore we see that
the transverse condition for our $(p-1)-$form field is naturally obtained upon
dimensional reduction. For a index not equal to $5$ we get 
\begin{equation}\label{transversepform}
(Re^{(\alpha_{p}+2)A}X^{\mu_{1}...\mu_{p-1}})'+ Re^{(\alpha_{p}+2)A}\partial_{\mu_{p}}X^{\mu_{1}...\mu_{p}}=0.
\end{equation}

First of all, we must split the field as done before by defining
$X_{T}^{\mu_{1}...\mu_{p}}\equiv X^{\mu_{1}...\mu_{p}}+\frac{(-1)^{p}}{\Box}\partial^{[\mu_{1}}\partial_{\nu_{1}}X^{\mu_{2}...\mu_{p}]\nu_{1}}$ and $X_{L}^{\mu_{1}...\mu_{p}}\equiv \frac{(-1)^{p-1}}{\Box}\partial^{[\mu_{1}}\partial_{\nu_{1}}X^{\mu_{2}...\mu_{p}]\nu_{1}}$. Observing now that
\begin{equation}
\partial_{\mu_{1}}Y^{\mu_{1}\mu_{2}...\mu_{p+1}}=\square X_{T}^{\mu_{2}...\mu_{p+1}};\;\;\;\;\partial_{\mu_{1}}Y^{\mu_{1}\mu_{2}...\mu_{p}}=\square X_{T}^{\mu_{2}...\mu_{p}},
\end{equation}
we see that the first term of Eq.  (\ref{pformnu}), just like in
the last section, is already decoupled from the longitudinal part.
However, the second term is not decoupled and if use the fact that
\begin{equation}
Y^{5\mu_{1}...\mu_{p}}=Y_{L}^{5\mu_{1}...\mu_{p}}+p! X_{T}^{'\mu_{1}...\mu_{p}}
\end{equation}
we can write the equation (\ref{pformnu}) as
\begin{eqnarray}
&& e^{\alpha_{p}A}\square X_{T}^{\mu_{1}...\mu_{p}}+(e^{\alpha_{p}A}\partial X_{T}^{\mu_{1}...\mu_{p}})' \nonumber \\
&&-\gamma_{p}Re^{(\alpha_{p}+2)A}X_{T}^{\mu_{2}...\mu_{p+1}}+\frac{1}{p!}(e^{\alpha_{p}A}Y_{L}^{5\mu_{1}...\mu_{p}})'\nonumber\\
&& -\gamma_{p}Re^{(\alpha_{p}+2)A}X_{L}^{\mu_{1}...\mu_{p}}=0 \label{eqXTXLpform},
\end{eqnarray}
and (\ref{pform5}) as
\begin{equation}\label{eqYLphipform}
\frac{1}{p!}\partial_{\mu_{1}}Y_{L}^{\mu_{1}\mu_{2}...\mu_{p}}-\gamma_{p}Re^{2A}X^{\mu_{2}...\mu_{p}}=0. 
\end{equation}

Therefore, we see clearly from Eq.  (\ref{eqXTXLpform}) that we have a coupling between
the transversal part of the $p-$form field, the longitudinal part and the $(p-1)-$form
field. From Eq.  (\ref{eqYLphipform}) we see that the $(p-1)-$form is coupled to the
longitudinal part of the $p-$form field. As in the case of the one form
field, we should expect that we have to uncouple the effective massive
equations for the gauge fields $X_{T}^{\mu_{1}\mu_{2}...\mu_{p}}$ and $X^{\mu_{2}...\mu_{p}}$
since both satisfy the transverse condition in four dimensions. Lets walk along
and prove this now. First of all note that using $\partial_{\mu_2}X^{\mu_{2}...\mu_{p}}=0$
we can show that 
\begin{equation}
 Y^{\mu_{1}...\mu_{p}}=\frac{(-1)^{p-1}}{\Box}\partial^{[\mu_{1}}\partial_{\nu}Y^{\mu_{2}...\mu_{p}]\nu}
\end{equation}
and we get an identity similar to that for the gauge field
\begin{equation}
Y_{L}^{\mu_{1}...\mu_{p}5}= p!\gamma_{p}\frac{Re^{2A}}{\Box}\partial^{[\mu_{1}}X^{\mu_{2}...\mu_{p}]},
\end{equation}
where in the last equation we have used equation (\ref{pform5}). Using now the transverse equation (\ref{transversepform}) we obtain
\begin{equation}
\left(e^{\alpha_{p}A}Y_{L}^{\mu_{1}...\mu_{p}5}\right)'=
p!\gamma_{p}Re^{(\alpha_{p}+2)A}X_{L}^{\mu_{1}...\mu_{p}} 
\end{equation}
and we get the equation of motion for the transversal part of $p$-form
\begin{equation}
 e^{\alpha_{p}A}\square X_{T}^{\mu_{1}...\mu_{p}}+(e^{\alpha_{p}A}\partial X_{T}^{\mu_{1}...\mu_{p}})' \nonumber -\gamma_{p}Re^{(\alpha_{p}+2)A}X_{T}^{\mu_{1}...\mu_{p}}=0.\label{XTpfull}
\end{equation}

Imposing now the separation of variables in the form $ X_{T}^{\mu_{1}...\mu_{p}}(z,x) = f(z)\tilde{X}_{T}^{\mu_{1}...\mu_{p}}(x)$ we obtain the mass equation
\begin{equation}
(e^{\alpha_{p}A}f')'-\gamma_{p}Re^{(\alpha_{p}+2)A}f =m_{X}^{2} p!e^{\alpha_{p}A}f, \label{eqfp}
\end{equation}
where the primes means derivative with respect to $z$.
Now, making $f(z) = e^{-\alpha_{p}A/2}\psi$ and using $e^{2A}R=-(D-1)[2A''+(D-2)A'^2]$, we can write the above equation in a Schr\"odinger form with potential given by
\begin{equation}\label{potp}
U(z)= [\frac{\alpha_{p}^{2}}{4}+(D-1)(D-2)\gamma_{p}]A'^{2}+[\frac{\alpha_{p}}{2}+2(D-1)\gamma_{p}]A''.
\end{equation}

The localized zero mode solution is given by $e^{pA}$ with $\gamma_{p}=[(D-2)-2\alpha_{p})/4(D-1)$.
For the $(p-1)$-form we have, imposing the separation of variables $X_{\mu_{2}...\mu_{p}}(z,x) = u(z)\tilde{X}_{\mu_{2}...\mu_{p}}(x)$ and from (\ref{pform5}) and (\ref{transversepform})  the mass equation
\begin{equation}
\left(Re^{-(\alpha_{p}+2)A}(Re^{(\alpha_{p}+2)A}u(z))'\right)'-\gamma_{p}Re^{2A}u(z)= m_{p-1}^{2}u(z).\label{equp}
\end{equation}
Just as in the last two section we see that we just have to use $u(z)=(Re^{(D-2p)A})^{1/2}\psi$ in (\ref{equp}) to get a Schr\"odinger equation with potential
\begin{eqnarray}\nonumber
U&=&\frac{1}{4}\left[(2\alpha_{p}+1)A'+(\ln R)'\right]^2\\ &-&\frac{1}{2}\left[(2\alpha_{p}+1)A''+(\ln R)''\right] +\gamma_{p}Re^{2A}.
\end{eqnarray}

From the above equation we see that we can recover all the previous cases. We also analyse the localizability of the field in a very simple way. For any metric which recovers the RS for large $z$ we get the asymptotic potential
\begin{equation}\label{pot(p-1)}
U(z)=\frac{1}{4}\left[(2\alpha_{p}+1)\right]^{2}A'^2-\frac{1}{2}\left[(2\alpha_{p}+1)A''\right] +\gamma_{p}Re^{2A}.
\end{equation}

The solution to the above equation is found by fixing $\gamma_{p}=(D+2+2\alpha_{p})/4(D-1)$.
Therefore we can see that the only case for localizing both fields happens for $p=(D-1)/2$. Now it is clear why for $D=5$ we have that KR field provides the localization of both fields. This is the result we want to stress here. This is possible due to the geometrical coupling and the field splitting described.

\section{Conclusions and Perspectives}

In this paper we have developed the idea that a geometrical coupling with the Ricci scalar can solve the problem of gauge field localization. We first showed that for any form field we can obtain decoupled equations of motion for the longitudinal and transverse components of the fields. We studied first the simplest cases, namely the Vector and Kalb-Ramond fields. From these we can understand how a generalization to $p-$forms can be obtained. Some points are worthwhile noting. First, we have found that for some specific value of coupling constant we can get the localization of any $p-$form. However, the $(p-1)-$form obtained by dimensional reduction can not be simultaneously localized. Despite of this, something very interesting happens in the Kalb-Ramond case in $D=5$. Here we get that through a dimensional reduction we naturally have the KR and the gauge field localized. This is a very important result since this gives a richer possibility of dynamics coming from a unique field in five dimensions. In fact, this can be seen as a new mechanism to localize the gauge vector field. As a byproduct It is also interesting to observe that for $p=0$ we get $\gamma_0=-(D-2)/4(D-1)$ what is exactly the conformal coupling to the scalar field. It remains to analyze other characteristics like resonant modes in this situation. The question about fermions with similar couplings can be interesting to another study. We can ask here, because of the fact of non-localization at the same time of fields coming from the procedure explained, if there is some physical criteria to choose one field or another. These are good questions to think about and are left to future works.   

\section*{Acknowledgment}

The authors would like to thanks the referee for the careful reading of the manuscript, suggesting modifications and corrections which improved it.   We also acknowledge the financial support provided by Funda\c c\~ao Cearense de Apoio ao Desenvolvimento Cient\'\i fico e Tecnol\'ogico (FUNCAP), the Conselho Nacional de 
Desenvolvimento Cient\'\i fico e Tecnol\'ogico (CNPq) and FUNCAP/CNPq/PRONEX.

\end{document}